\documentclass[conference,compsoc]{IEEEtran}
\ifCLASSOPTIONcompsoc
  \usepackage[nocompress]{cite}
\else
  \usepackage{cite}
\fi
\ifCLASSINFOpdf
\else
\fi
\usepackage{amsmath,amssymb,amsfonts}
\usepackage{graphicx,color}
\usepackage{textcomp}
\usepackage{algorithm}
\usepackage{algpseudocode}
\usepackage{caption}
\usepackage[labelfont=bf,textfont=md]{caption}
\usepackage{array}
\usepackage{subcaption}
\usepackage{soul}
\usepackage{float}

\begin{document}
%
\title{StoX-Net: Stochastic Processing of Partial Sums for Efficient In-Memory Computing DNN Accelerators}



%


\author{\IEEEauthorblockN{Ethan G. Rogers\IEEEauthorrefmark{1}\IEEEauthorrefmark{3},
Sohan Salahuddin Mugdho\IEEEauthorrefmark{1}\IEEEauthorrefmark{3},
Kshemal K. Gupte\IEEEauthorrefmark{1} and 
Cheng Wang\IEEEauthorrefmark{1}\IEEEauthorrefmark{2}}
\IEEEauthorblockA{\IEEEauthorrefmark{1}Department of Electrical and Computer Engineering\\
Iowa State University of Science and Technology,
Ames, IA 50010}
\IEEEauthorblockA{\IEEEauthorrefmark{2}Corresponding Author: chengw@iastate.edu}
\IEEEauthorblockA{\IEEEauthorrefmark{3}Both authors contributed equally to this work}
}



\maketitle

\begin{abstract}
Crossbar-based in-memory computing (IMC) has emerged as a promising platform for hardware acceleration of deep neural networks (DNNs). 
However, the energy and latency of IMC systems are dominated by the large overhead of the peripheral analog-to-digital converters (ADCs).
To address such ADC bottleneck, here we propose to implement stochastic processing of array-level partial sums (PS) for efficient IMC. Leveraging the probabilistic switching of spin-orbit torque magnetic tunnel junctions, the proposed PS processing eliminates the costly ADC and achieves significant improvement in energy and area efficiency. 
To mitigate accuracy loss, we develop PS-quantization-aware training that enables backward propagation across stochastic PS. Furthermore, a novel scheme with an inhomogeneous sampling length of the stochastic conversion is proposed. 
When running ResNet20  on the CIFAR-10 dataset, our architecture-to-algorithm co-design demonstrates up to 16x, 8x, and 10x improvement in energy, latency, and area, respectively, compared to IMC with standard ADC. Our optimized design configuration using inhomogeneous sampling of stochastic PS achieves 130x (24x) improvement in Energy-Delay-Product compared to IMC with full precision ADC (sparse low-bit ADC), while maintaining near-software accuracy at various benchmark classification tasks.
\end{abstract}


%
\IEEEpeerreviewmaketitle

\section{Introduction}
\begin{figure}[h]
  \centering
  \includegraphics[width=\linewidth]{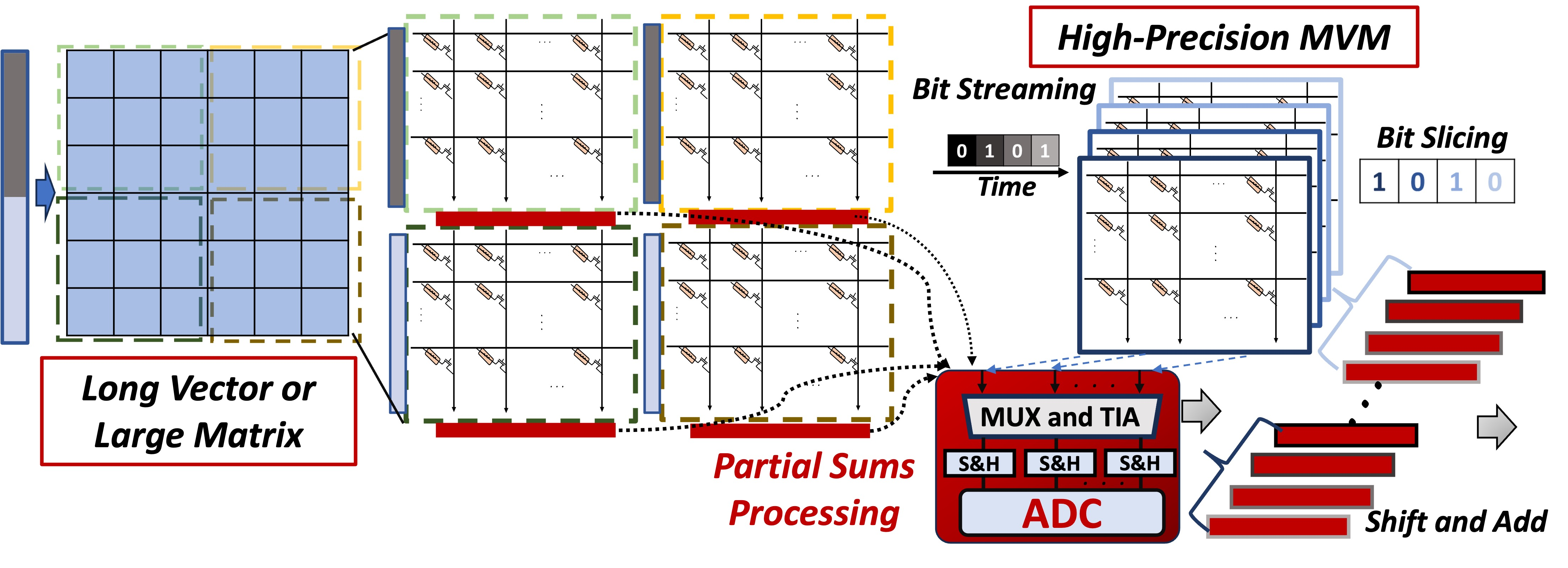}
  \vspace{-15pt}
  \caption{ADC bottleneck incurred at the partial sum processing in  IMC crossbar architecture.}
  \label{fig: partial_sum}
  \vspace{-18pt}
\end{figure}

\IEEEPARstart{C}{rossbar-based} analog in-memory computing (IMC) has demonstrated great potential for Deep Neural Network (DNN) acceleration by achieving high efficiency and massive parallelism at processing matrix-vector-multiplications (MVMs)\cite{isaac}, which dominates most state-of-the-art DNN workloads\cite{ibm_gemms}. However, the hardware efficiency of IMC sees a severe bottleneck due to the significant overhead of peripheral analog-to-digital converters (ADCs), which are required to ensure robust data communication in a large-scale system (Fig.\ref{fig: partial_sum}). ADC is found to consume over 60-80$\%$ of the energy and chip area of IMC hardware \cite{crossbar_proceeding, isaac}. 
Particularly, the large area overhead of high-precision A-D conversion requires one ADC to be shared by multiple columns in a crossbar. Such design leads to a severe throughput bottleneck since the processing of the array-level outputs across multiple columns has to be done sequentially.

While various recent works have aimed to mitigate the ADC bottleneck \cite{Sparse_ReRAM, mixed_quantization, ADC-Less}, minimizing ADC overhead in IMC while still maintaining satisfactory inference accuracy remains a significant challenge. 
Various factors contribute to the difficulty in addressing the ADC bottleneck. First, since the array-level partial sums (PS) are not represented in DNN algorithms, the standard quantization of activation and weights can not directly address the ADC precision for PS. Second, quantization-aware training with PS quantization desires backward propagation with incorporation of array-level variables, which is challenging as it requires re-designing the computational graph of the DNN model. Third, the required ADC bit precision for MVM may vary significantly depending on both the algorithmic attributes (such as sparsity and DNN layer dimension) and hardware attributes (such as array size and bits per memory cell). Accommodating these varying scenarios requires reconfigurability and flexibility in ADC, leading to increased overhead and design complexity.

In this work, we propose \textbf{stochastic processing} of the array-level partial sum for efficient IMC by leveraging a spin-orbit-torque magnetic tunnel junction (SOT-MTJ) with simple digital circuitry. 
To address the accuracy drop due to the stochastic PS, a hardware-aware training methodology is developed. Notably, by exploiting the error tolerance of neural networks, our hardware-software co-design achieves the conversion of analog crossbar currents to digital representations with a small number of temporal samplings (1-4 bits for most cases). Such findings \textbf{address the long-standing issues of large latency} incurred from using stochastic bit sequences in computation. 
Moreover, we conducted extensive system-level hardware simulation to validate that the low-overhead crossbar peripheral based on the proposed spintronic devices/circuits demonstrates significant improvement in energy efficiency and area savings. 
Our proposed IMC design eliminates the ADC bottleneck and opens up an exciting direction where stochastic computation could play a vital role in designing efficient non-von Neumann DNN accelerators. Our major contributions are the following:
\begin{itemize}
  \item We propose StoX-Net, an IMC architecture with aggressively quantized array-level partial sums based on stochastic switching dynamics of SOT-MTJs with simple CMOS peripherals. The ADC bottleneck in IMC is eliminated, leading to significant improvement in hardware efficiency.
  \item We develop a comprehensive framework of PS quantization-aware training that addresses the accuracy degradation due to the stochastic conversion. In particular, both the device-level stochastic MTJ switching behavior, and key architectural attributes including bit slicing and array size, are incorporated into backward propagation. 
  Our model with 1-bit stochastic PS achieves state-of-the-art accuracy at benchmarking image classification tasks.
  \item We identify that the state-of-the-art quantization of convolutional DNN has severe limitations due to the \textbf{first convolution layer remaining at high precision}. 
  To address this challenge, we develop layer-wise inhomogeneous sampling numbers based on a Monte-Carlo sensitivity analysis to enable aggressive PS quantization on the first convolution layer. As a result, accuracy loss due to quantization is mitigated with minimal additional overhead, demonstrating an advantageous trade-off between efficiency and accuracy.
\end{itemize}
\vspace{-10pt}
\section{Background}
\subsection{DNN Acceleration With IMC}
Analog IMC with the standard weight-stationary configuration accelerates DNN inference by minimizing the data movement of MVM processing, alleviating the von Neumann memory bottleneck. 
To process a large-scale DNN workload, large matrices and MVMs with high-precision operands need to be partitioned into multiple crossbar arrays. To ensure robust data communication, an analog crossbar IMC macro is connected with other components through digital-to-analog converters (DAC) at the input and ADC at the crossbar output. 
The resolution of DAC at each row is typically designed at 1 bit for area saving, 
while high-precision input activation will be converted to bit streams over multiple time steps (bit streaming). 
As for high-precision weights, due to the technological limitation in the number of bits per cell, a full-precision weight matrix will be partitioned into several slices of sub-arrays (bit slicing).

The required ADC solution N for array-level PS processing is $N = log_2(N_{row}) + I + W - 2$
, where N\textsubscript{row} is the number of activated rows; I is input bits per stream; and W is bits per slice.
Since the energy, area, and latency overhead of ADC increase significantly with bit precision \cite{adc_survey}, ADC becomes the bottleneck of efficiency and performance in IMC design.
\vspace{-5pt}
\subsection{Stochastic Spintronics For MVM in DNN}
Several emerging NVM technologies, including resistive memory (ReRAM) and magnetic random access memory (MRAM),
have been explored for machine learning acceleration. 
Recent explorations have demonstrated that spintronic devices exhibit sufficient endurance to realize both synaptic weight storage and neural activation \cite{mram_endurance}. 
Particularly, an SOT-MTJ under an excitation current with varying magnitude will have the magnetization switching probability as a sigmoidal function of the current, which provides a direct emulation of a stochastic neuron \cite{stochastic_neuron} with significantly higher area/energy efficiency compared to CMOS implementations. 
Moreover, the separate read/write paths of SOT-MTJs offer immense design flexibility by eliminating the constraint of write current density\cite{Sharma2021}. All-spin neuro-synaptic processors based on stochastic switching of SOT-MTJ have been proposed \cite{all-spin_stochastic}. 
We will explore SOT-MTJs for efficient PS processing of the crossbar output.

\subsection{Related Works}
Various works have investigated co-designing the IMC hardware architecture and DNN algorithms to reduce the ADC precision. 
One major thrust focuses on quantization with hardware-aware training. In \cite{BNN_batch_norm, BNN_resnet} binary neural networks exhibit improved efficiency for IMC, but the models are limited to 1-bit representation. 
Recent works \cite{ADC-Less, BitSplit-Net, EPSQ} implemented bit slicing/streaming to map workloads with multi-bit weights/activations. Aggressively reducing the ADC precision to 1-bit will essentially enable using sense amplifiers. 
However, 1-bit PS showed sizeable accuracy degradation even after hardware-aware re-training \cite{BitSplit-Net, EPSQ}. Moreover, the state-of-the-art quantization-aware training keeps \textbf{the first convolution layer at full precision} \cite{IR-Net, ADC-Less, Bi-real}. However, the compute-intensive first convolution layer in image processing can dominate the overall computation workloads. Keeping a full-precision first layer severely limits the overall improvement in energy and throughput. 

Another major thrust is to exploit sparsity through pruning and re-training to enable lower ADC precision \cite{Samba, Sparse_ReRAM}. Higher sparsity reduces the range of possible values of MVM output and thus reduces the ADC resolution requirement. 
As a result, sparsity-aware IMC with reconfigurable ADC resolution can improve energy and latency.
However, 
it is important to note that the \textit{area} of the peripheral ADC circuitry remains large in order to handle the \textit{highest} bit precision of the reconfigurable design. Such large area overhead still hinders the hardware parallelism in IMC architecture. 

 \begin{figure*}[h]
  \centering
  \includegraphics[width=\linewidth]{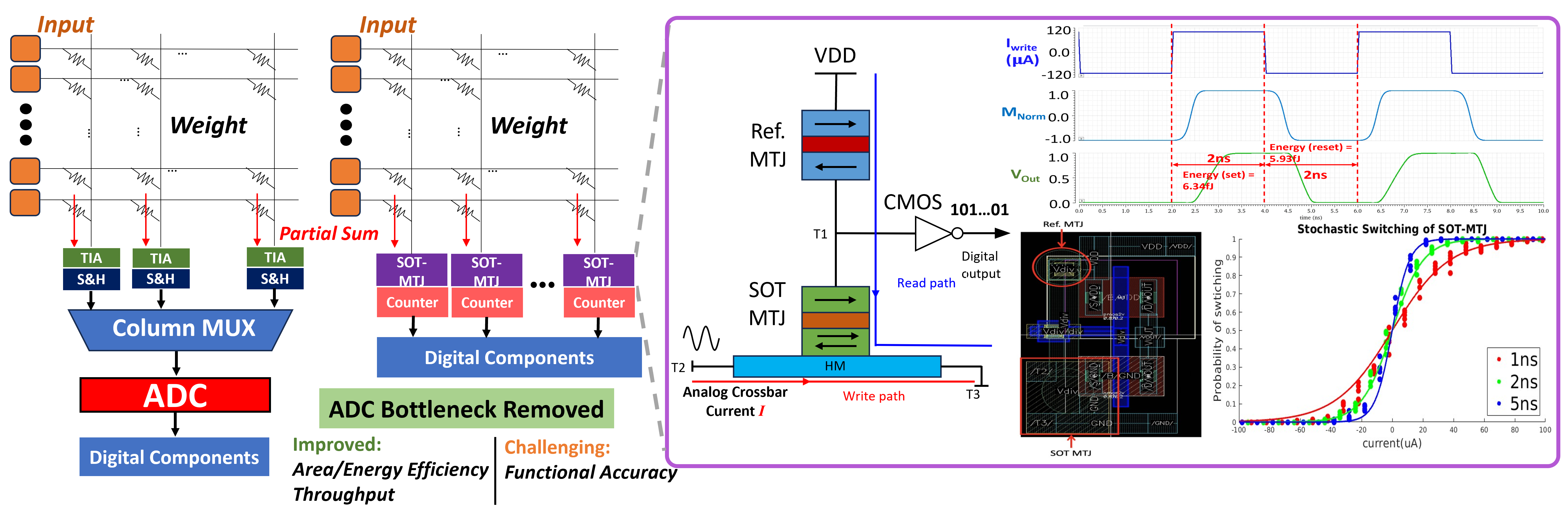}
  \vspace{-15pt}
  \caption{The overview of the proposed crossbar MVM processing with stochastic MTJ converter.}
  \label{fig: overview}
  \vspace{-18pt}
\end{figure*} 

Compared to related works, our proposed StoX-Net exhibits distinctive characteristics. First, we, for the first time, represent the array-level PS by hardware-inspired stochastic bits, and incorporate the stochastic representation into training with bit slicing. Second, we leverage the sampling of stochastic bits to provide reconfigurability in the conversion precision without sacrificing the area efficiency and hardware parallelism.
Third, to achieve better scalability to larger-scale input, we venture into aggressive PS quantization of the first convolution layer and demonstrate a balanced performance of accuracy and hardware efficiency.

\section{In Memory Computing with Stochastic Partial Sums}\label{sec:Method}
The proposed Stox-Net is built on IMC crossbar arrays where current-voltage converters, column-shared MUXs, and peripheral ADCs are replaced by a row of stochastic SOT-MTJs (Fig. \ref{fig: overview}). 
We first present the device/circuit design of our stochastic PS processing components and then implement stochasticity-aware training of quantized DNNs with bit slicing considered \cite{Bi-real, IR-Net}. 
\vspace{-10pt}
\subsection{Stochastic PS Processing with MTJ}
         
The stochastic processing based on the current-driven probabilistic switching of SOT-MTJs is shown in Fig. \ref{fig: overview}. A voltage divider circuit combined with a CMOS inverter functions as an analog-to-stochastic converter using MTJs with tunnel magnetoresistance ratio (TMR) of 350-450\% \cite{Sharma2021}.
The switching probability is simulated using a MATLAB-based macro-spin Landau Lifshitz Gilbert (LLG) equation simulator with the SOT current included. 
We verify the functionality of the stochastic MTJ-converter by simulating the voltage-divider circuit with MTJs as illustrated in Fig. \ref{fig: overview}. A modified version of Spintronics Interdisciplinary Center (Spinlib) model \cite{spinlib} is used to model the SOT-MTJ switching. 
We emulate the crossbar output current using a current source that provides both the positive and negative currents that would accumulate in the crossbar columns (in the range of $-100 \mu A$ to $100 \mu A$). Note that, in practical operation, the crossbar current is driven by the crossbar input voltages supplied by the digital-to-analog converter (DAC). The terminal T3 of the SOT-MTJ would have a supply voltage that is half of the input supply voltage so that the voltage difference is bi-polar across the heavy metal (HM) layer of the SOT-MTJ, allowing bi-directional current to pass through the HM layer. We demonstrate the operation of the MTJ converter via the waveforms in Fig. \ref{fig: overview}. 
\begin{table}[h]
    \centering
    \begin{tabular}{|l|c|}
        \hline
        \textbf{Parameter} & \textbf{Value}\\
        \hline
        SOT-MTJ dimension & 90nm x 70nm x 2.5nm\\
        \hline
        $R_{LRS}$ & 57 k$\Omega$\\
        \hline
        TMR & 4.4\\
        \hline
        $t_{ox}$ & 1.3 nm\\
        \hline
        HM resistivity ($\rho$) & 160 $\mu\Omega$cm \cite{Sharma2021} \\
        \hline
        HM dimensions & 144nm x 112nm x 3.5nm\\
        \hline
        $I_{write}$ & 0 - $\pm$100$\mu$A\\
        \hline
        Supply Voltage & 1 V\\
        \hline
        Ref. MTJ resistance & 140 k$\Omega$\\
        \hline
    \end{tabular}
    \caption{Device parameters}
    \label{tab:device-parameters}
\end{table}

Our simulations demonstrate a reliable conversion latency of 2ns, indicating rapid response times suitable for high-frequency applications. Energy consumption was measured at $6.35 fJ$ for the set operation and at $5.94 fJ$ for the reset operation, indicating an average energy consumption of $6.14 fJ$ per conversion. 
The stochastic MTJ converter's area depicted in Fig. \ref{fig: overview} is 0.9108 $\mu$$m^2$ based on Global Foundries 22FDSOI technology PDK \cite{GF22FDX}. We scale the area to 28 nm technology node for our study. The dimensions of the reference MTJ are taken as per the GF 22FDX Technology Design Manual, while the dimensions of the SOT MTJ are obtained from Table \ref{tab:device-parameters}. Our proposed SOT-MTJ-based converter is orders of magnitude more energy and area-efficient than ADC, leading to significant improvement of system-level hardware efficiency compared to standard IMC.
\vspace{-10pt}
\subsection{Hardware-aware Training of Quantized DNN}
\subsubsection{Crossbar Mapping with Stochastic PS}
We map large DNN workloads into crossbars with finite array size and limited bits per cell with bit slicing from \cite{isaac} and input/weight representation from \cite{ADC-Less}. 
Following Algorithm \ref{alg:cap}, mapping a convolution layer of kernel size \(K_h\) $*$ \(K_w\) and \(C_{in}\) input channels using a crossbar array of \(R_{arr}\) rows results in \(N_{arrs}\) PS arrays, where \(N_{arrs}=ceil(\frac{m}{R_{arr}})= ceil(\frac{K_h*K_w*C_{in}}{R_{arr}})\). All input and weight values are then quantized (\(A_q,W_q\)) and split into subarrays (\(A_q^{[0, N_{arrs})},W_q^{[0, N_{arrs})}\)). Weight bits are further split into slices physically mapped to different subarrays, and the input bits are temporally streamed into the crossbar. 
Each steamed input bit on each sliced subarray  will contribute to one PS, leading to \(A_s*W_s\) partial output that needs to be converted to digital domain and then shift-and-added.
SOT-MTJs are connected to process these array-level output currents with the option of multiple samplings. The output of the PS aggregation is then divided by $N_{arrs}*N_{samples}$ for normalization to [-1, 1] as layer-wise MVM output.

\vspace{2pt}
\begin{algorithm}[h]
\caption{Crossbar MVM with Stochastic Partial Sum}\label{alg:cap}
\begin{algorithmic}
\State \textbf{Input: }  \(A_{l} \in [-1, 1], A_b, A_s \in [1, A_b], W_l \in \Re, W_b, W_s \in [1, W_b],\) \(R_{arr} =\) max PS array column length, \(m =\) input vector size 
\State \textbf{Output: } Hardware-aware \(O_{l}\)
\Procedure {StoX MVM} {}
    \State $N_{arrs} \gets ceil(\frac{m}{R_{arr}})$  \Comment{\# of PS arrays}
    \State $A_q \gets Q(A_l, A_b)$ \Comment{Quantize \(W_l, A_l\)}
    \State $W_q \gets Q(bn(W_l), W_b)$ \Comment{bn($W_l$) column-wise}
    \State $A_q^{[0, N_{arrs})} \gets Split(A_q, N_{arrs}, A_s)$ \Comment{Make PS arrays}
    \State $W_q^{[0, N_{arrs})} \gets Split(W_q, N_{arrs}, Q_s)$
    \For{$i \in [0, N_{arrs})$}
        \State $O^{i}_q \gets MVM(W_q^i, \vec{A_q^i})$ \Comment{Perform MVMs}
        \For{\texttt{$j \in [0, N_{samples})$}} \Comment{Multi-sampling}
            \State $O^{i,j}_{MTJ} \gets MTJ(O^{i}_q)$ \Comment{MTJ conversion}
            \State $O^{i}_{MTJ} \gets O^{i}_{MTJ} + O^{i,j}_{MTJ}$
            \Comment{Samples counter}
            
        \EndFor
        \State $O^{i}_{S\&A} \gets S\&A(O^{i}_{MTJ})$ 
        \Comment{Shift \& Add slices}
        \State $O_l \gets O_l + O^{i}_{S\&A}$
    \EndFor
    \State $O_l \gets \frac{O_l}{N_{arrs}*N_{samples}}$
    \Comment{Normalize output to [-1, 1]}
\EndProcedure
\end{algorithmic}
\end{algorithm}
 With the binary states of SOT-MTJ encoded as (-1,1), the switching probability versus input current is emulated by the \textit{tanh} function: 
\begin{equation} \label{eq2}
MTJ(x) = \begin{cases} 
      -1 & tanh(\alpha x) < rand \\
      1 & tanh(\alpha x)\geq rand \\
    \end{cases}
\end{equation}
where $\alpha$ is the sensitivity parameter. Increasing $\alpha$ makes the tanh curve more step-like, approaching a deterministic 1-bit sense amplifier (1b-SA). It has been shown that encoding binary activation to (-1,1) provides better representation capability than encoding to (0,1) \cite{XOR_Net}.  
The MTJ's effective sensitivity in hardware can be altered by tuning the range of crossbar current when mapping MVM operations to hardware.


\begin{figure}[ht]
    \centering
    \includegraphics[width=0.99\linewidth]{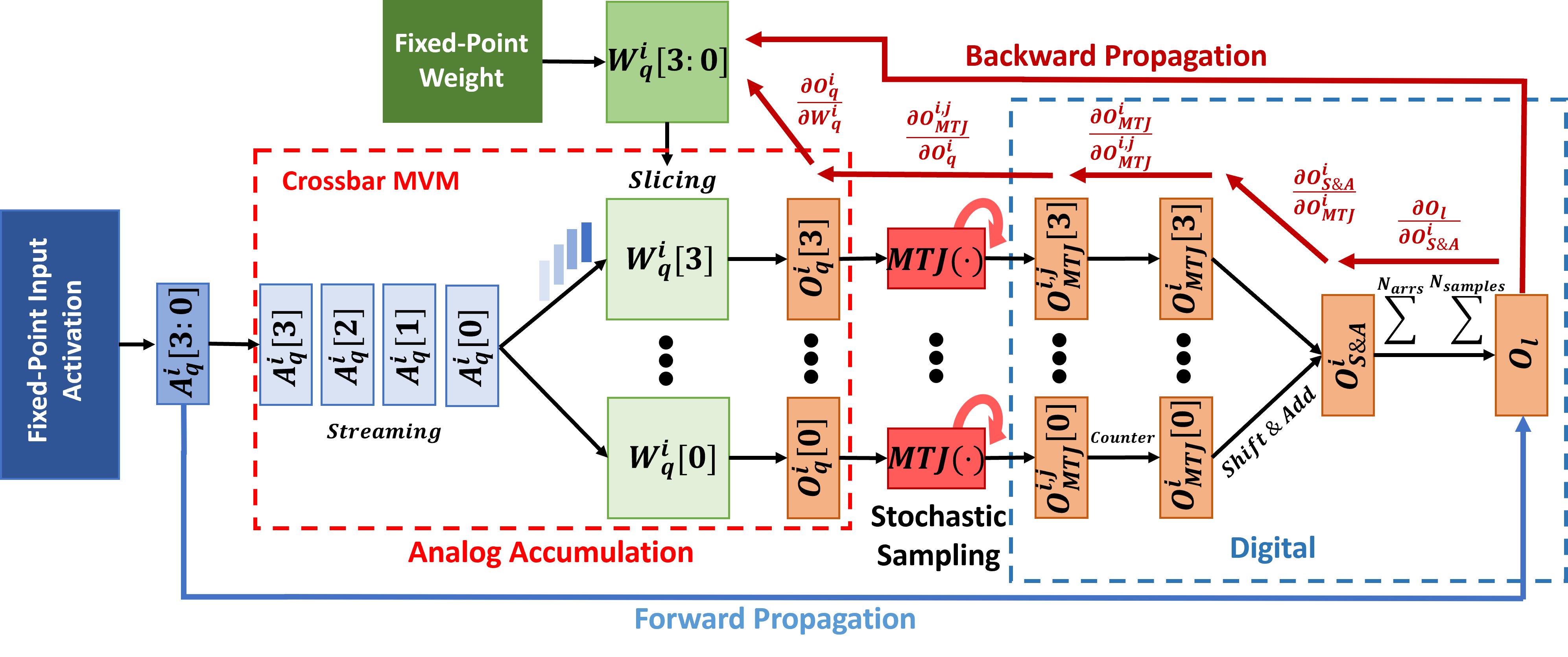}
    \vspace{-10pt}
    \caption{Computational flow of MVM operation in the proposed StoX-Net architecture. In this example, inputs and weights are 4-bit fixed-point values ($A_b = W_b$ = 4). Slices and streams are 1-bit ($A_s = W_s = 1$).}
    \label{fig:high_level_prop}
\end{figure}
\subsubsection{Backpropagation with Stochastic binarization}
During backward propagation, the loss $\frac{\partial L}{\partial W_l}$ across a single convolutional layer with partial sums follows the high-level representation in Fig. \ref{fig:high_level_prop} and is shown in-depth via the following equations:
\vspace{-5pt}
\begin{equation}
    \frac{\partial L}{\partial W_l} = 
    \frac{\partial L}{\partial O_l}
    \sum_{i=1}^{N_{arrs}}
    \frac{\partial O_l}{\partial W_{l}}[i].
\end{equation}
Where $O_l$ is the layer's final output summing all the array-level outputs after the shift-and-add operation ($S\&A$). 
For each layer output vector that goes through $S\&A$ and sampling through the MTJ converter, the gradient is calculated as follows:
\begin{equation}
    \frac{\partial O_l}{\partial W_l}[i] = 
    \frac{\partial O_l}{\partial O_{S\&A}^{i}}[i] *
    \frac{\partial O_{S\&A}^{i}}{\partial O^{i}_{MTJ}}*
    \sum_{j=1}^{N_{samples}}
    \frac{\partial O^{i}_{MTJ}}{\partial W_l}
\end{equation}


Each array $i$'s gradient backtracks to the shift-and-added (S\&A) result of the stochastic MTJ output, which counters (sums) the output over $j$ MTJ samples. We approximate the MTJ's gradient using a straight-through estimator (STE), clamping values outside its saturation range to avoid exploding gradients. The MTJ can then be connected to the layer's weights via:

\begin{equation}
    \frac{\partial O^{i}_{MTJ}}{\partial W_{l}}[j]=
    \frac{\partial O^{i}_{MTJ}}{\partial O^{i,j}_{MTJ}}[j]*
    \frac{\partial O^{i,j}_{MTJ}}{\partial O^{i}_q}
    \frac{\partial O^{i}_q}{\partial W_q^i}
    \frac{\partial W^{i}_q}{\partial W_{bn}}
    \frac{\partial W_{bn}}{\partial W_l}.
\end{equation}
Where weight quantization, array-splitting, and normalization algorithms are considered with respect to the layer's weights. For each subarray $i$, the layer output is the sum of all $S\&A$ terms. As a result, for a given $m$-long input stream and $n$ weight slices, $m*n$ terms would exist. In the final MVM output, these PS terms are scaled (through shifting) differently based on their $m$ and $n$ indices, such that a set of scalars $\{\frac{2^{(m*n-1)}}{2^{m*n}-1},\dots,\frac{1}{{2^{m*n}-1}}\}$ can be defined. Each of these scalar terms is applied to an MTJ-sampled PS, making the partial derivative a set of scalars. Our MTJ in backpropagation is treated as a straight-through estimator (STE), and the MVM output $O_q^i$ of each PS array results from well-defined linear operations. 
The weight quantization's gradients are similar to that of \cite{IR-Net}. The gradient of batch normalization operation is also a scalar. 
Considering all of the operations involved in generating the MVM output, the gradient representation for backward propagation can be reduced to:
\begin{equation}
    \frac{\partial L}{\partial W_l} = 
    \frac{\partial L}{\partial O_l}
    \frac{\partial O_l}{\partial \sum O_{arrs}^{MTJs}}
    \frac{\partial \sum O_{arrs}^{MTJs}}{\partial W_q}
    \frac{\partial W_q}{\partial W_{bn}}
    \frac{\partial W_{bn}}{\partial W_l}.
\end{equation}

\begin{figure}[ht]
    \centering    \includegraphics[width=0.89\linewidth, height=3.5cm]{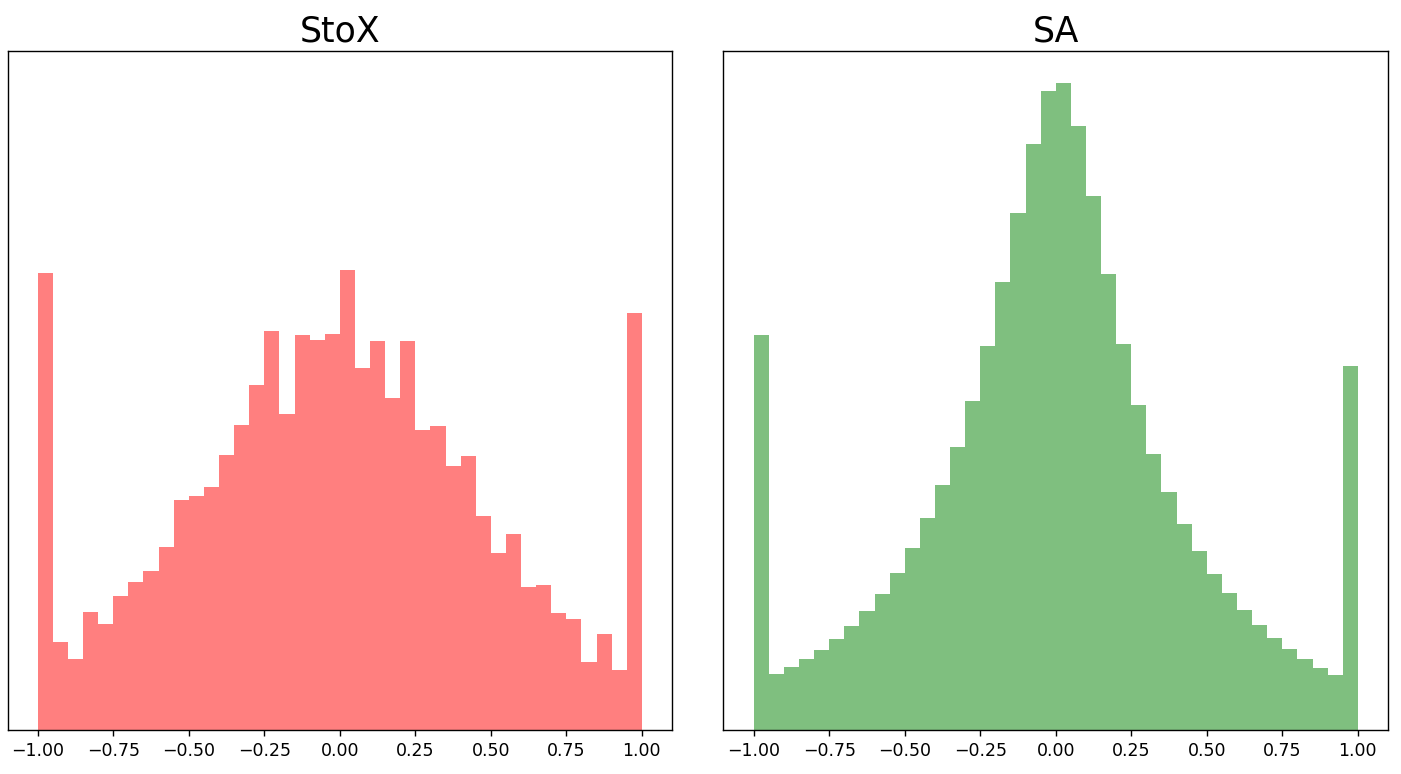}
    \caption{The distribution of normalized array-level MVM outputs collected in a DNN model trained with stochastic MTJs (``StoX'') compared with a model with sense amplifier (SA) behavior.}
    \label{fig:mtj_dist}
    \vspace{-10pt}
\end{figure}
Fig. \ref{fig:mtj_dist} compares the data distribution of crossbar output using the stochasticity-aware trained network (StoX) and the model trained with deterministic 1b PS (SA). We observe that incorporating stochasticity in training generates a broader and less polarized distribution with more intermediate values. Conceptually, such a distribution will facilitate using more of the sloped region of the \textit{tanh} curve of switching probability. 
In contrast, the training with deterministic 1-bit SA will encourage the PS output to be concentrated around 0 and $\pm$1. 

\begin{figure}[h]
    \centering
    \includegraphics[width=0.85\linewidth]{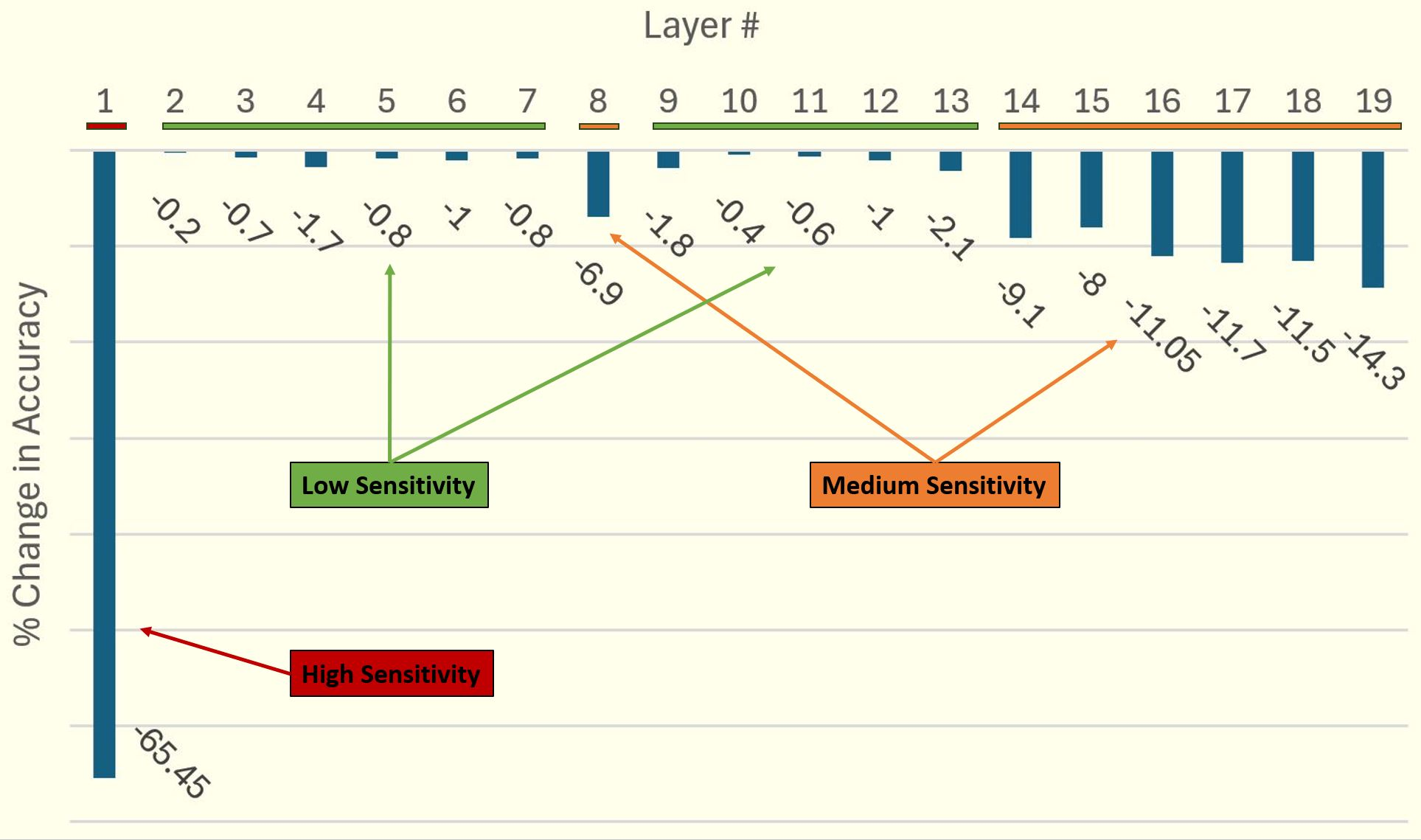}
    \caption{Monte Carlo simulation for determining the layer-wise importance and error sensitivity. }
    \vspace{-10pt}
    \label{fig:monte_carlo}
\end{figure}
\subsubsection{Error Mitigation with Multi-sampling}To mitigate the errors due to stochastic bit generation, 
we propose multi-sampling to recover accuracy with a trade-off of hardware efficiency (latency and energy). 
Counting from more SOT-MTJ samples will lead to a multi-level representation of the analog output.
We explored up to 8 samples per conversion since the total conversion energy and latency increase linearly with increasing number of samples. 
To only increase the number of samplings for the layers that are most sensitive to PS precision, we perform a Monte Carlo simulation on our network's trainable layers to guide an \textbf{inhomogeneous multi-sampling} scheme. For each layer, we apply a uniform random perturbation to its weights at inference and then measure the significance of the weights of that layer to the model accuracy. Fig. \ref{fig:monte_carlo} shows the first layer being most susceptible to perturbation, indicating the most significance, while layers close to the output classifications are only moderately susceptible to error.
Based on this analysis, we implement a model with mixed layer-wise sampling numbers (``Mix''), where layers with higher (lower) sensitivity are given more (fewer) MTJ samples. We achieve an accuracy similar to a homogeneous 4-sample network with only a small increase in operations compared to the 1-sample network (see discussions on the result in Section \ref{sec:hardware-efficiency}). 

\section{Experimental Results: Functional Accuracy and Hardware Efficiency}\label{sec:results}
\subsection{Evaluation Methodology}
We evaluate both the functional accuracy and hardware efficiency of StoX-Net designs based on implementing convolutional neural networks for image classification on the MNIST and CIFAR10 datasets. Our evaluations include both a quantized first convolution layer (QF) and a high-precision first convolution layer (HPF). All subsequent layers in the evaluated models adopt the proposed MTJ-based stochastic conversion. All QF models take 8 samples per MTJ conversion in the first layer due to the importance of Layer 1 as shown by the Monte Carlo analysis in Fig. \ref{fig:monte_carlo}. 
When deterministic 1-bit sense amplifiers (SA) are used for PS processing, we label it as ``HPF+1b-SA".

We denote X-bit weight, Y-bit activation, and Z-bits per slice as $XwYaZb_s$. For example, a software-level model with 4-bit activation and 4-bit weight mapped to 2-bits per slice is $4w4a2b_s$. The baseline  \textit{StoX} network has $4w4a4b_s$, $\alpha=4$, $R_{arr}=256$, and 1 sample per MTJ conversion is applied using HPF configuration. The naming of the evaluated models highlights their differences from this baseline. The integer in front of the abbreviated letters (such as QF and HPF) means the number of MTJ sampling for all the other layers except the first layer \textit{conv-1}. For example, 1-QF (4-QF) uses 1 (4) stochastic sample(s) of the MTJ converters in all layers except the quantized first layer, which still requires 8 samples per MTJ conversion. Mix-QF uses mixed sampling rates based on the Monte Carlo simulation with an 8-sample quantized first layer. 

\begin{figure*}[h]
    \centering
    \includegraphics[width=0.96\textwidth, height=3.5cm]{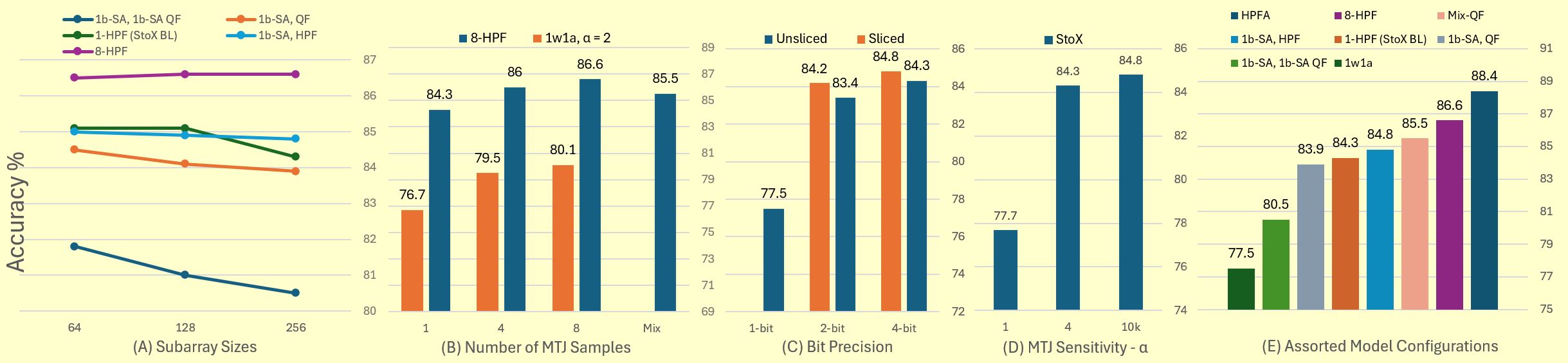}
    \caption{Accuracy across various configurations and an ablation study of StoX-Net's performance on CIFAR-10. Networks are only described by their differences from the baseline StoX model described in Section \ref{sec:results}. 
}
    \label{fig:main_ablation}
    \vspace{-15pt}
\end{figure*}

As the PS processing of the first layer and later layers can be set separately (see discussions on Fig.\ref{fig:main_ablation}),``1b-SA, 1b-SA QF'' means that step-like 1b-SA (\textit{tanh} with high $\alpha$) is applied to \textit{all} layers (including the first conv layer), showing the lowest accuracy due to severe quantization loss. ``1b-SA, QF'' applies stochastic MTJ with 8 samples to the first layer while 1b-SA is applied to all other layers. 
``1b-SA, HPF'' is our version of the 1b deterministic ``HPF+1b-SA'' in the reference. 
We also include a model with binarized weight and activation (1w1a) to evaluate the performance of our approach on extremely quantized models.
``Unsliced'' refers to the case where the number of bits per memory device in hardware is the same as the number of bits in a weight element, while ``Sliced'' has 1 bit per memory cell.

\begin{figure}[h]
\vspace{-10pt}
  \includegraphics[width=.95\linewidth]{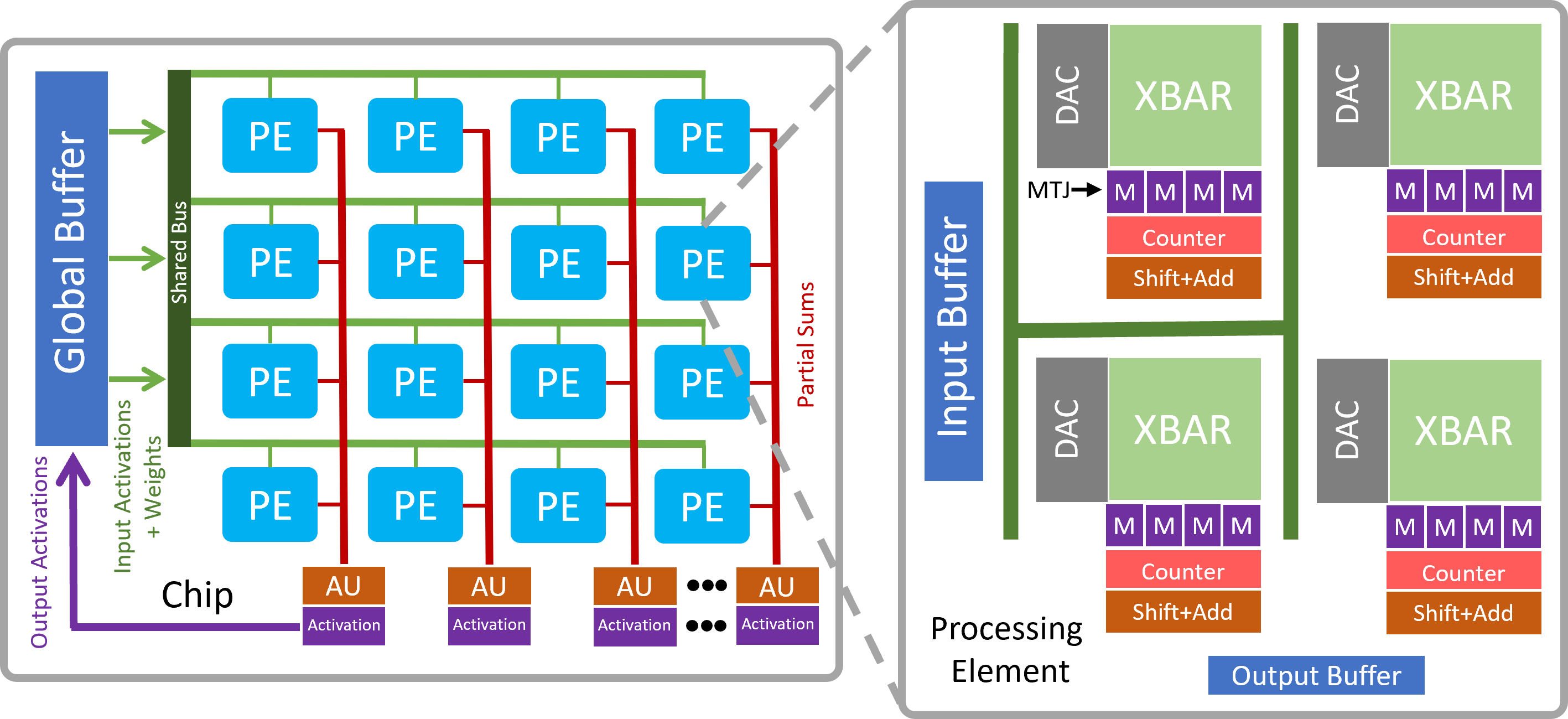}
  \vspace{-3pt}
  \caption{Evaluated In-Memory Computing System}
  \label{fig:sim-architecture}
\end{figure}

We train ResNet-20 models based on the proposed convolutional methodology in Section \ref{sec:Method}. For CIFAR-10, we train with hyperparameters similar to \cite{IR-Net}, and for MNIST we lower epochs from 400 to 25. We note that larger models and more channels (such as such as VGG-9 or ResNet-18) can achieve higher accuracy for the evaluated benchmark tasks. However, our development focuses on ResNet-20, which has a compact model size and a small number of channels. We compare our StoX-Net results with similar implementations from the literature.

Moreover, we compare the hardware efficiency of the implementation of StoX-Net architecture with a standard IMC, with standard high-precision ADC for all layers (HPFA). Moreover, a sparse configuration (SFA) with reduced (full precision - 1)-bit ADC is also included as a stronger baseline for comparison.  
We evaluate hardware efficiency by simulating a standard IMC architecture similar to that of ISAAC \cite{isaac}, as shown in Fig. \ref{fig:sim-architecture}. We model and evaluate our architecture using Accelergy/Timeloop \cite{timeloop,accelergy,accelergy-pim}. Our baseline HPFA architecture results were verified to be consistent with previous works \cite{design-space}\cite{puma}.
Signed weights are represented using 2 cells per weight as in \cite{ADC-Less}, leading to the possibility of having both positive and negative currents in the column bit lines.
We use Successive Approximation Register (SAR) ADC with the area and power metrics scaled for 28nm technology based on comprehensive surveys and analysis of ADC architectures \cite{adc_survey} \cite{design-space,neuromorphic-processors}. 
Table \ref{tab:component-table} summarizes key parameters for the hardware components. Because ADCs dominate the resources of IMC architectures in terms of energy, latency, and area \cite{adcdominance}, we focus on the impact of replacing the ADC with the stochastic MTJ-converter. 
\begin{table}[ht]
    \centering
    \vspace{-8pt}
    \begin{tabular}{|l|c|c|}
        \hline
        \textbf{Component} & \textbf{Energy/Action} & \textbf{Area/instance}\\
         & \textbf{($pJ$)} & \textbf{($\mu m^2$)}\\
        \hline
        DAC \cite{timeloop} & 2.99e-2 & 0.127\\
        \hline
        Xbar Cell(1b/2b) \cite{puma} & (6.16e-3/4.16e-3) & 0.0308\\
        \hline
        ADC(FP/Sparse) \cite{adc_survey} & (2.137/1.171) & (6600/2700)\\
        \hline
        MTJ-Converter & 6.14e-3 & 1.47\\
        \hline
    \end{tabular}
    \caption{Energy and Area of Simulated Hardware Components}
    \vspace{-12pt}
    \label{tab:component-table}
\end{table}
\begin{table}[ht]
    \centering
    \vspace{-5pt}
    \begin{tabular}{c|c|c|c|c|}
        \cline{2-5}
        & \multicolumn{3}{c|}{StoX-Net} & \multicolumn{1}{c|}{HPF+1b-SA \cite{ADC-Less}} \\\hline
        \multicolumn{1}{|c|}{\textbf{Samples}} & 1-QF & 4-QF & Mix-QF & \multicolumn{1}{c|}{/}\\\hline
        \multicolumn{1}{|c|}{\textbf{1w1a1\(b_s\)}} & 96.8 & 97.6 & 97.2 & 98.9 \\\hline
        \multicolumn{1}{|c|}{\textbf{2w2a2\(b_s\)}} & 97.7 & 98.3 & 97.9 & - \\\hline
        \multicolumn{1}{|c|}{\textbf{2w2a1\(b_s\)}} & 98.0 & 98.4 & 98.2 & 99.0 \\\hline
        \multicolumn{1}{|c|}{\textbf{4w4a4\(b_s\)}} & 97.9 & 98.5 & 98.1 & 99.2 \\\hline
        \multicolumn{1}{|c|}{\textbf{4w4a1\(b_s\)}} & 98.2 & 98.4 & 98.3 & - \\\hline
    \end{tabular}
    \caption{StoX-Net implementation of a modified ResNet-20 on MNIST Handwritten Digits with $R_{arr}=128$.}
    \label{tab:MNIST_table}
\end{table}
\begin{table}[ht]
    \centering
    \vspace{-10pt}
    \newcolumntype{?}{!{\vrule width 1.5pt}}
    \begin{tabular}{c|c|c|c|c|c|c|}
        \cline{2-7}
        & \multicolumn{4}{c|}{StoX-Net} & \multicolumn{2}{c|}{HPF+Quantized}\\\hline
        \multicolumn{1}{|c|}{\textbf{Samples}} & 1& 4& 8 & Mix & \multicolumn{2}{c|}{/ }\\\hline
        \multicolumn{1}{|c|}{\textbf{QF}} & 76.1& \textbf{83.3}& \textbf{83.8}& \textbf{80.6}& -- & --\\\hline
        \multicolumn{1}{|c|}{\textbf{HPF}} & 84.3& \textbf{86.0}& \textbf{86.6}& \textbf{85.5}& 84.2 \cite{ADC-Less} & 85.4 \cite{IR-Net}\\\hline

    \end{tabular}
    \caption{StoX \(4w4a4b_s\) ResNet-20 on CIFAR-10 with $R_{arr}=256$.}
    \label{tab:CIFAR10_table}
    \vspace{-20pt}
\end{table}
\subsection{Functional Accuracy}\label{sec:accuracy}
The proposed StoX-Net can reach within 1\% of the reference HPF model's accuracy for both MNIST and CIFAR-10 datasets, as summarized in Tables \ref{tab:MNIST_table} \& \ref{tab:CIFAR10_table}.
Fig. \ref{fig:main_ablation} presents the ablation study of our algorithmic approach, The impacts of various techniques are summarized in panel (E).
Both (A) and (E) panels of Fig. \ref{fig:main_ablation}  show that applying 1b PS quantization to all layers (``1b-SA, 1b-SA QF'') suffers a significant loss, which is consistent with the state of the art that requires a high-precision first layer. Replacing 1b-SA with a multi-sample stochastic MTJ improves the accuracy, and using the largest sampling numbers achieves the highest accuracy. 
By enabling an 8-sample stochastic \textit{conv-1} layer only, ``1b-SA, QF'' experiences immediate (over 3.4\%) accuracy improvement over ``1b-SA, 1b-SA QF''. 
Among the HPF configurations, while ``1b-SA, HPF'' is slightly better than the 1-HPF StoX baseline, more samplings per MTJ clearly improve the accuracy of StoX model. Such observations suggest that multi-sampling in stochastic PS processing is an effective knob to enhance the accuracy of implementing IMC with aggressive PS quantization.

Fig. \ref{fig:main_ablation} (A) also shows that large arrays suffer accuracy degradation as more information loss occurs when PS quantization is applied to an increased range of possible crossbar output values. 
Fig. \ref{fig:main_ablation}(B) demonstrates that \textbf{stochastic multi-sampling consistently mitigates accuracy loss}.
In Fig. \ref{fig:main_ablation}(C), bit slicing helps mitigate the accuracy degradation of quantization. 
Fig. \ref{fig:main_ablation} (D) shows that for models with one sample, a smaller $\alpha$ (a more sloped \textit{tanh} curve) is more erroneous due to stochasticity-induced fluctuation, while high $\alpha$ (step-like) is more deterministic and less noisy. 
However, we re-iterate that with a sloped \textit{tanh} curve (small $\alpha$), better accuracy can be obtained through multi-sampling (Table \ref{tab:CIFAR10_table}). Such accuracy improvement is not possible with deterministic 1b-SA. 
Interestingly, 
the proposed ``Mixed-HPF'' manages to only increase the number of MVM conversions by 14.3\% compared to the 1-sample model, but reaches 85.5\% accuracy (similar to the 4-sample network's performance). As discussed in the following section, a small increase of samples per MTJ conversion achieves the best trade-off in terms of achieving high hardware efficiency and maintaining accuracy.

\begin{figure}[h]
\includegraphics[width=.95\linewidth,height=4cm]{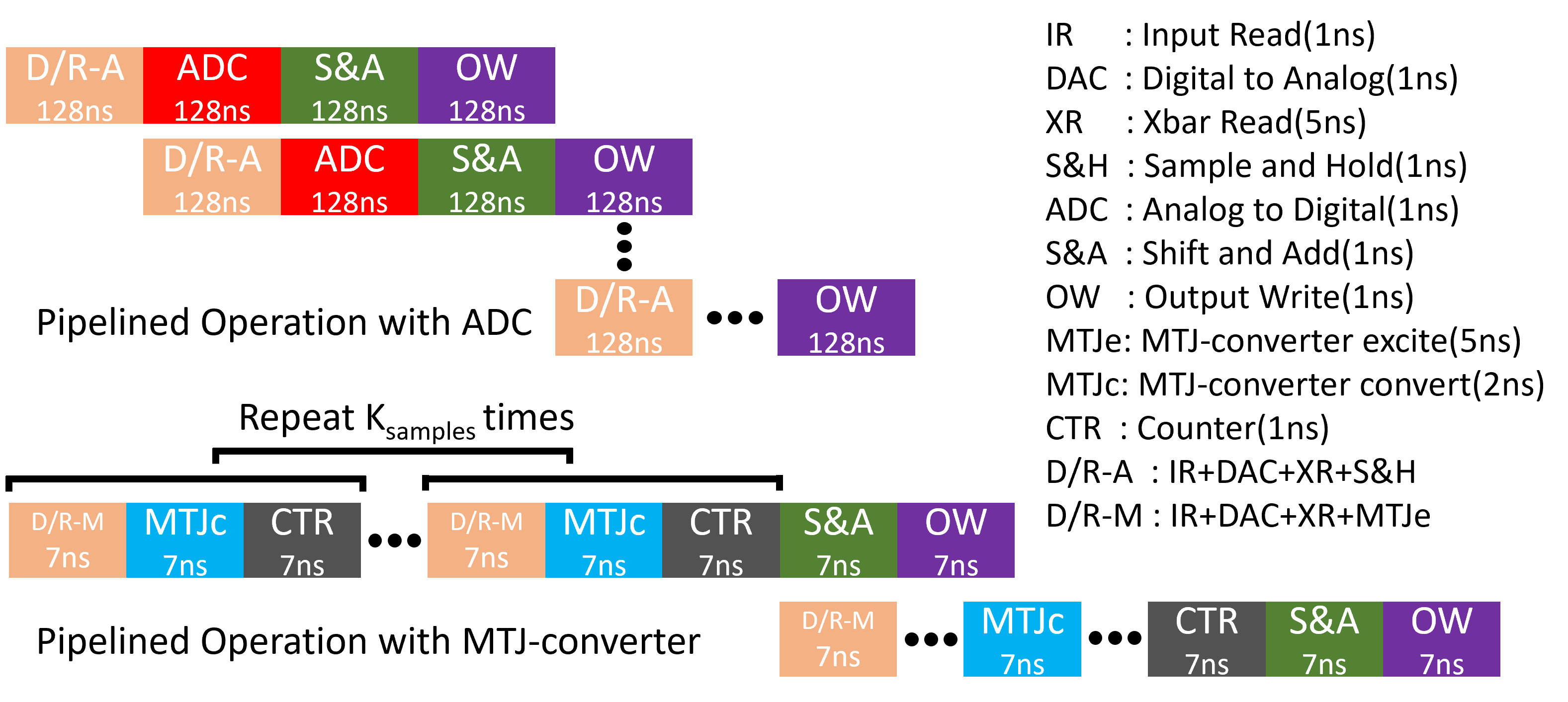}
  \vspace{-5pt}
  \caption{Pipelined MVM operation in a crossbar IMC using ADC versus using MTJ-converter}
  \vspace{-10pt}
  \label{fig:pipeline}
\end{figure}
\subsection{Hardware Efficiency}\label{sec:hardware-efficiency}
\vspace{-5pt}
\begin{figure*}[ht]
  \centering
  \begin{subfigure}[t]{0.79\textwidth}
      \includegraphics[width=\textwidth]{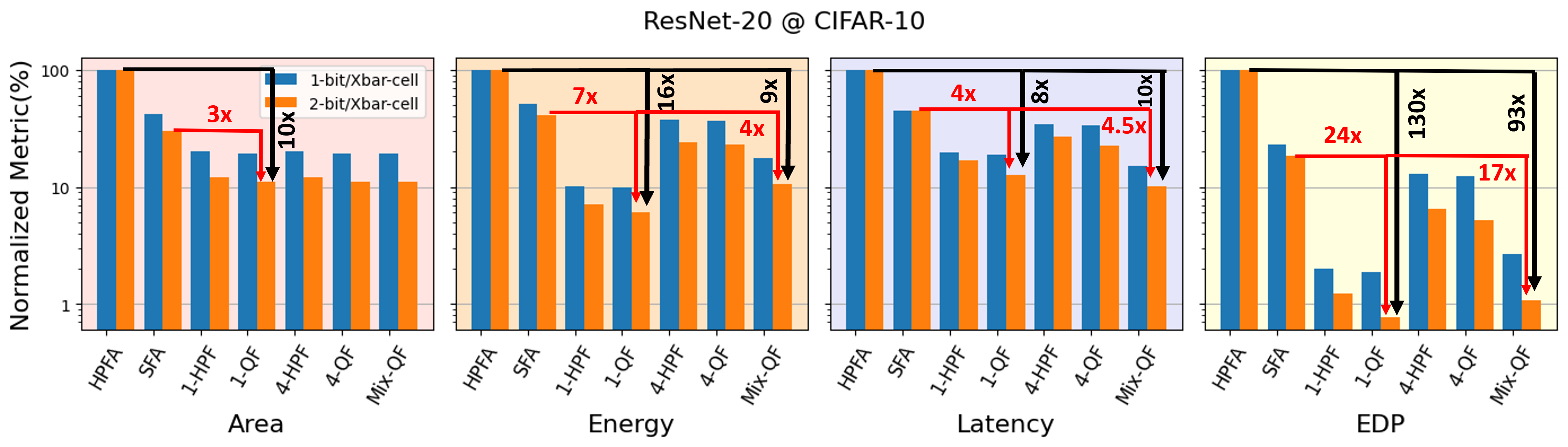}
      \vspace{-15pt}
      \caption{ }
      \label{fig:comparison-resnet20}
  \end{subfigure}
  \begin{subfigure}[t]{0.202\linewidth}
      \includegraphics[width=\linewidth]{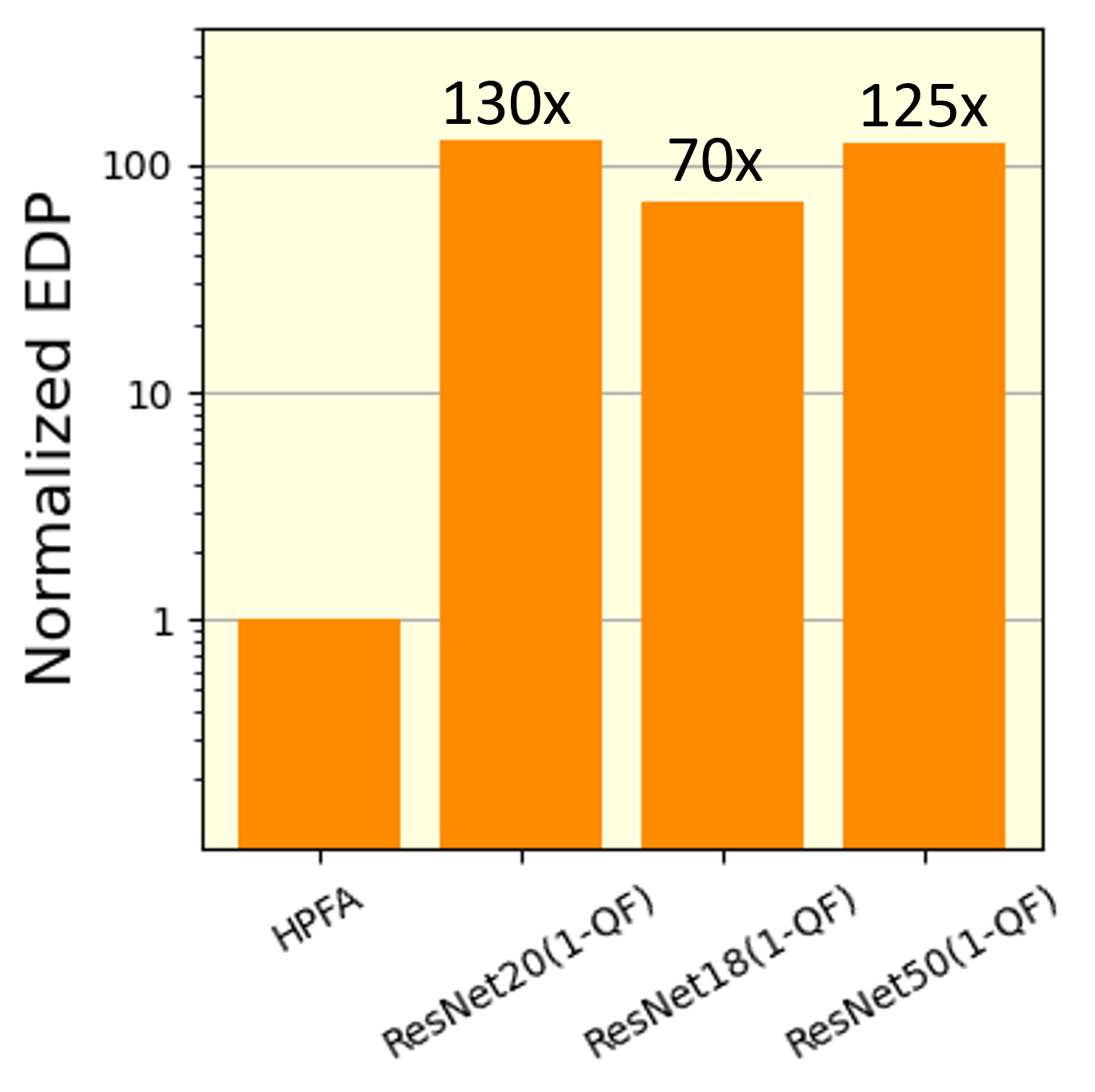}
      \vspace{-15pt}
      \caption{ }
      \label{fig:threemodel}
  \end{subfigure}
  \vspace{-5pt}
  \caption{Hardware efficiency evaluated by simulation of IMC architecture using Timeloop \cite{timeloop}. (\ref{fig:comparison-resnet20}) Normalized metrics between the baseline configurations and optimized configurations running ResNet-20 on CIFAR-10. (\ref{fig:threemodel}) EDP improvement of ResNet-20 on CIFAR-10, ResNet-18 and ResNet-50 on Tiny ImageNet compared to respective baseline (HPFA).}
  \vspace{-10pt}
  \label{fig:comparison}
\end{figure*}
The proposed area-efficient and highly parallel PS processing directly impacts the datapath design pattern of crossbar MVM operation. Fig. \ref{fig:pipeline} shows the pipelined operations inside a crossbar using an ADC (top) and a stochastic MTJ converter (bottom). 
In standard IMC, the length of each pipeline stage is determined by the longest stage, i.e. the ADC readout of all shared columns in a crossbar. 
Our optimization \textbf{alleviates the throughput bottleneck} by parallelizing all columns with compact MTJ converters, thereby considerably reducing the length of the pipeline stage. 


Our hardware performance evaluation is summarized in Fig. \ref{fig:comparison}. Full precision ADC (HPFA) and Sparse design (SFA) are our baselines for comparison. The first observation is that SOT-MTJ converters significantly enhance parallelism with reduced area overhead, showing large savings in both area and EDP. Additionally, our design utilizes weights and activations of reduced precision, contributing further to efficiency.
Processing reduced precision models with our proposed hardware design achieved improvements of area, energy, and latency by up to $10$x, $16$x, and $8$x, respectively. Our design achieves up to $24$x ($130$x) EDP improvement compared to an IMC with sparse (full-precision) ADC. 
 
We further evaluate the trade-off of efficiency and accuracy by varying the number of stochastic samplings. Multi-sampling improves accuracy compared to 1-QF, as observed in Fig.\ref{fig:main_ablation}, at the cost of reduced throughput and more energy consumption. Searching for the optimal trade-off needs to minimize the increase of samplings while achieving high inference accuracy. By exploiting the Monte-Carlo-based error sensitivity analysis shown in Fig. \ref{fig:monte_carlo}, 
we managed to only implement 2 or 4 samplings to a few layers in the deep neural network while maintaining the majority of PS processing at 1 sampling. As a result, such Mix-QF models only slightly increase the total number of MTJ conversions compared to the 1-sample models, maintaining relatively high hardware efficiency ($17-93$x EDP gain) while achieving accuracy close to the 4-sample network (see Table \ref{tab:CIFAR10_table}).
 We further demonstrate in Fig. \ref{fig:threemodel} that our architecture can be scaled to larger models and workloads while maintaining remarkable efficiency improvement. 

\vspace{-10pt}
\section{Conclusion}
\vspace{-10pt}
We develop StoX-Net, an efficient IMC design that eliminates the ADC bottleneck by processing array-level partial sum using stochastic MTJ. 
Based on our hardware-software co-design with stochasticity-aware training, DNN inference accuracy reaches close to the state-of-the-art accuracy while achieving up to 10x area efficiency and 24-130x improvement in hardware efficiency characterized by EDP. Moreover, we demonstrate that leveraging the multi-sampling capability of stochastic PS can quantize the first convolution layer (within $2\%$ accuracy loss compared to other quantized models), which is not possible in prior quantization-aware training. 
Future work will explore applying stochastic architectures to larger DNNs and datasets. More sophisticated designs of DNN architecture beyond standard networks (such as ResNet) can be created using automated machine learning such as network architecture search. 
Our work suggests exciting opportunities for utilizing stochastic computation and spintronics to develop next-generation AI hardware accelerators. 
\vspace{-10pt}
\bibliographystyle{IEEEtran}
\bibliography{reference}

\begin{thebibliography}{10}
\providecommand{\url}[1]{#1}
\csname url@samestyle\endcsname
\providecommand{\newblock}{\relax}
\providecommand{\bibinfo}[2]{#2}
\providecommand{\BIBentrySTDinterwordspacing}{\spaceskip=0pt\relax}
\providecommand{\BIBentryALTinterwordstretchfactor}{4}
\providecommand{\BIBentryALTinterwordspacing}{\spaceskip=\fontdimen2\font plus
\BIBentryALTinterwordstretchfactor\fontdimen3\font minus \fontdimen4\font\relax}
\providecommand{\BIBforeignlanguage}[2]{{%
\expandafter\ifx\csname l@#1\endcsname\relax
\typeout{** WARNING: IEEEtran.bst: No hyphenation pattern has been}%
\typeout{** loaded for the language `#1'. Using the pattern for}%
\typeout{** the default language instead.}%
\else
\language=\csname l@#1\endcsname
\fi
#2}}
\providecommand{\BIBdecl}{\relax}
\BIBdecl

\bibitem{isaac}
A.~Shafiee, A.~Nag, N.~Muralimanohar, R.~Balasubramonian, J.~P. Strachan, M.~Hu, R.~S. Williams, and V.~Srikumar, ``Isaac: A convolutional neural network accelerator with in-situ analog arithmetic in crossbars,'' \emph{ACM SIGARCH Comp. Arch. News}, 2016.

\bibitem{ibm_gemms}
B.~Fleischer \emph{et~al.}, ``Unlocking the promise of approximate computing for on-chip ai acceleration,'' IBM Research Blog, 2020.

\bibitem{crossbar_proceeding}
I.~Chakraborty, M.~Ali, A.~Ankit, S.~Jain, S.~Roy, S.~Sridharan, A.~Agrawal, A.~Raghunathan, and K.~Roy, ``Resistive crossbars as approximate hardware building blocks for machine learning: Opportunities and challenges,'' \emph{Proceedings of the IEEE}, 2020.

\bibitem{Sparse_ReRAM}
T.-H. Yang, H.-Y. Cheng, C.-L. Yang, I.-C. Tseng, H.-W. Hu, H.-S. Chang, and H.-P. Li, ``Sparse reram engine: Joint exploration of activation and weight sparsity in compressed neural networks,'' in \emph{ISCA}, 2019, pp. 236--249.

\bibitem{mixed_quantization}
S.~Huang, A.~Ankit, P.~Silveira, R.~Antunes, S.~R. Chalamalasetti, I.~El~Hajj, D.~E. Kim, G.~Aguiar, P.~Bruel, S.~Serebryakov \emph{et~al.}, ``Mixed precision quantization for reram-based dnn inference accelerators,'' in \emph{ASP-DAC}, 2021, pp. 372--377.

\bibitem{ADC-Less}
U.~Saxena, I.~Chakraborty, and K.~Roy, ``Towards adc-less compute-in-memory accelerators for energy efficient deep learning,'' in \emph{2022 DATE}, 2022, pp. 624--627.

\bibitem{adc_survey}
B.~Murmann, ``{ADC Performance Survey 1997-2023},'' [Online]. Available: \url{https://github.com/bmurmann/ADC-survey}.

\bibitem{mram_endurance}
J.~J. Kan, C.~Park, C.~Ching, J.~Ahn, Y.~Xie, M.~Pakala, and S.~H. Kang, ``A study on practically unlimited endurance of stt-mram,'' \emph{IEEE Transactions on Electron Devices}, vol.~64, no.~9, pp. 3639--3646, 2017.

\bibitem{stochastic_neuron}
A.~Sengupta, P.~Panda, P.~Wijesinghe, Y.~Kim, and K.~Roy, ``Magnetic tunnel junction mimics stochastic cortical spiking neurons,'' \emph{Scientific reports}, vol.~6, no.~1, p. 30039, 2016.

\bibitem{Sharma2021}
T.~Sharma, C.~Wang, A.~Agrawal, and K.~Roy, ``Enabling robust sot-mtj crossbars for machine learning using sparsity-aware device-circuit co-design,'' in \emph{IEEE/ACM ISLPED}, 2021.

\bibitem{all-spin_stochastic}
G.~Srinivasan, A.~Sengupta, and K.~Roy, ``Magnetic tunnel junction enabled all-spin stochastic spiking neural network,'' in \emph{Design, Automation \& Test in Europe Conference \& Exhibition (DATE), 2017}.\hskip 1em plus 0.5em minus 0.4em\relax IEEE, 2017, pp. 530--535.

\bibitem{BNN_batch_norm}
Y.~Kim, H.~Kim, and J.-J. Kim, ``Neural network-hardware co-design for scalable rram-based bnn accelerators,'' \emph{ArXiv}, vol. abs/1811.02187, 2018.

\bibitem{BNN_resnet}
Y.~Kim, H.~Kim, J.~Park, H.~Oh, and J.-J. Kim, ``Mapping binary resnets on computing-in-memory hardware with low-bit adcs,'' in \emph{2021 DATE}, 2021, pp. 856--861.

\bibitem{BitSplit-Net}
H.~Kim, Y.~Kim, S.~Ryu, and J.-J. Kim, ``Algorithm/hardware co-design for in-memory neural network computing with minimal peripheral circuit overhead,'' in \emph{2020 57th ACM/IEEE DAC}.\hskip 1em plus 0.5em minus 0.4em\relax IEEE, 2020, pp. 1--6.

\bibitem{EPSQ}
Y.~Kim, H.~Kim, and J.-J. Kim, ``Extreme partial-sum quantization for analog computing-in-memory neural network accelerators,'' \emph{ACM JETC}, vol.~18, no.~4, pp. 1--19, 2022.

\bibitem{IR-Net}
H.~Qin, R.~Gong, X.~Liu, M.~Shen, Z.~Wei, F.~Yu, and J.~Song, ``Forward and backward information retention for accurate binary neural networks,'' in \emph{2020 IEEE/CVF CVPR}, 2020, pp. 2247--2256.

\bibitem{Bi-real}
Z.~Liu, B.~Wu, W.~Luo, X.~Yang, W.~Liu, and K.-T. Cheng, ``Bi-real net: Enhancing the performance of 1-bit cnns with improved representational capability and advanced training algorithm,'' in \emph{ECCV}, 2018, pp. 722--737.

\bibitem{Samba}
D.~E. Kim, A.~Ankit, C.~Wang, and K.~Roy, ``Samba: Sparsity aware in-memory computing based machine learning accelerator,'' \emph{IEEE Transactions on Computers}, 2023.

\bibitem{spinlib}
\BIBentryALTinterwordspacing
Z.~Wang, W.~Zhao, E.~Deng, J.-O. Klein, and C.~Chappert, \emph{STT SOT MTJ}, 2015. [Online]. Available: \url{http://www.spinlib.com/STT\_SOT\_MTJ.html}
\BIBentrySTDinterwordspacing

\bibitem{GF22FDX}
\BIBentryALTinterwordspacing
GlobalFoundries, \emph{22FDX®-EXT Technology Design Manual Rev. 1.0\_4.1}, 2023. [Online]. Available: \url{https://gf.com/technology-platforms/fdx-fd-soi/}
\BIBentrySTDinterwordspacing

\bibitem{XOR_Net}
S.~Zhu, L.~H.~K. Duong, and W.~Liu, ``Xor-net: An efficient computation pipeline for binary neural network inference on edge devices,'' in \emph{2020 IEEE ICPADS}, 2020, pp. 124--131.

\bibitem{timeloop}
A.~Parashar, P.~Raina, Y.~S. Shao, Y.-H. Chen, V.~A. Ying, A.~Mukkara, R.~Venkatesan, B.~Khailany, S.~W. Keckler, and J.~Emer, ``Timeloop: A systematic approach to dnn accelerator evaluation,'' in \emph{2019 IEEE ISPASS}, 2019, pp. 304--315.

\bibitem{accelergy}
Y.~N. Wu, J.~S. Emer, and V.~Sze, ``Accelergy: An architecture-level energy estimation methodology for accelerator designs,'' in \emph{2019 IEEE/ACM ICCAD}, 2019, pp. 1--8.

\bibitem{accelergy-pim}
Y.~N. Wu, V.~Sze, and J.~S. Emer, ``An architecture-level energy and area estimator for processing-in-memory accelerator designs,'' in \emph{2020 IEEE ISPASS}, 2020, pp. 116--118.

\bibitem{design-space}
K.~He, I.~Chakraborty, C.~Wang, and K.~Roy, ``Design space and memory technology co-exploration for in-memory computing based machine learning accelerators,'' in \emph{IEEE/ACM ICCAD}, ser. ICCAD '22, 2022.

\bibitem{puma}
A.~Ankit, I.~E. Hajj, S.~R. Chalamalasetti, G.~Ndu, M.~Foltin, R.~S. Williams, P.~Faraboschi, W.-m.~W. Hwu, J.~P. Strachan, K.~Roy \emph{et~al.}, ``Puma: A programmable ultra-efficient memristor-based accelerator for machine learning inference,'' in \emph{24th ASPLOS}, 2019, pp. 715--731.

\bibitem{neuromorphic-processors}
Q.~Wang, Y.~Kim, and P.~Li, ``Neuromorphic processors with memristive synapses: Synaptic interface and architectural exploration,'' \emph{ACM Journal on Emerging Technologies in Computing Systems (JETC)}, vol.~12, no.~4, pp. 1--22, 2016.

\bibitem{adcdominance}
W.~Haensch, A.~Raghunathan, K.~Roy, B.~Chakrabarti, C.~M. Phatak, C.~Wang, and S.~Guha, ``Compute in-memory with non-volatile elements for neural networks: A review from a co-design perspective,'' \emph{Advanced Materials}, vol.~35, no.~37, p. 2204944, 2023.

\end{thebibliography}
\vspace{-10pt}
\end{document}